\documentclass[journal]{IEEEtran}
\usepackage[utf8]{inputenc}

\usepackage[english]{babel}

\usepackage[T1]{fontenc}

\usepackage[displaymath,switch]{lineno}

\usepackage{microtype}
\usepackage{siunitx}
\usepackage{graphicx}

\usepackage{amsmath}
\usepackage{mathtools}

\usepackage[capitalise]{cleveref}
\Crefname{figure}{Fig.}{Figs.}
\Crefname{equation}{Equation}{Equation}
\crefname{equation}{}{}

\usepackage{todonotes}

\usepackage{booktabs}
\usepackage{multirow}

\usepackage[mode=image]{standalone}

\usepackage{tikz}
\usetikzlibrary{calc,arrows,arrows.meta,decorations.markings,positioning,shadows,shapes.geometric}%

\usepackage{pgfplots}
\pgfplotsset{width=8cm,compat=1.14}

\usepackage{bytefield}

\usepackage{textgreek}%

\usepackage{bold-extra}

\tikzstyle{busconn} = [thick, decoration={markings,mark=at position
	1 with {\arrow[scale=1, thick]{open triangle 60}}},
	double distance=4pt, shorten >= 5.7pt,
	preaction = {decorate},
	postaction = {draw,line width=4pt,white,shorten >= 5.3pt}]
\tikzstyle{innerWhite} = [semithick, white,line width=1.4pt, shorten >= 4.5pt]

\DeclareSIUnit\bps{bps}%
\DeclareSIUnit\Sa{S}%

\newcommand{\avgbr}[1]{\langle#1\rangle}

\ifdefined\LINENUMBERS
\newcommand{\revlabel}[1]{\label{#1}\linelabel{#1}}
\else
\newcommand{\revlabel}[1]{\label{#1}}
\fi

\begin{document}

\bstctlcite{IEEEexample:BSTcontrol}

\title{DPTC---an FPGA-based Trace Compression}

\author{
  G.~Bruni %
  and~H.~T.~Johansson%
\thanks{G. Bruni and H.~T.~Johansson
    are with the Department of Physics, Chalmers University of Technology,
    SE-412 96 G{\"o}teborg, Sweden.}%
\thanks{This work was supported by the Swedish Research Council,
  the Scientific Council for Natural and Engineering Sciences
  under grant 2017-03839 and
  the Council for Research infrastructure
  under grant 822-2014-6644.}%
}

\IEEEpubid{0000--0000/00\$00.00}

\maketitle

\ifdefined\LINENUMBERS
\linenumbers
\fi

\begin{abstract}
  Recording of flash-ADC traces is challenging from both the transmission bandwidth and storage cost perspectives.
This work presents a configuration-free lossless compression algorithm, which addresses both limitations, by compressing the data on-the-fly in the controlling FPGA. Thus it can easily be used directly in front-end electronics.
The method first computes the differences between consecutive samples in the traces, thereby concentrating the most probable values around zero.
The values are then stored as groups of four, with only the necessary least-significant bits in a variable-length code, packed in a stream of 32-bit words.
To evaluate the efficiency, the storage cost of compressed traces is modeled as a baseline cost including ADC noise, and a cost for pulses that depends on amplitude and width.
The free parameters and the validity of the model are determined by compressing artificial traces with varying characteristics.
The compression method was also applied to actual data from different types of detectors.
A typical storage cost is around 4 to 5 bits per sample.
Code for the FPGA implementation in VHDL and for the CPU decompression routine in C are available as open source software, both able to operate at speeds of 400 Msamples/s.

\end{abstract}

\begin{IEEEkeywords}
Analog-to-digital conversion (ADC), data acquisition, data compression, field programmable gate array (FPGA), front-end electronics, lossless compression, real-time data acquisition, open source, variable-length code, VHDL.
\end{IEEEkeywords}

\IEEEpeerreviewmaketitle

\section{Introduction}

\IEEEPARstart{T}{his} work is motivated by developments in data handling in nuclear and particle physics.
However, its applicability is not limited to those fields.
Experiments in nuclear and particle physics are growing, which implies an increasing amount of data that needs to be handled.
This is caused by an increase in the number of detectors employed, finer segmentation and higher event rates.
Of particular interest for this work is the recording of signal traces, because this is associated with a dramatic increase of data that need to be transferred, compared with a simple digitization of pulse amplitudes.

\revlabel{revtest}%
To illustrate the development of experimental setups,
we consider two front-line particle physics experiments almost
30 years apart.
We compare the ATLAS (\emph{A Toroidal LHC ApparatuS}) experiment at LHC, CERN, which started data-taking in 2009, with the UA1 (\emph{Underground Area 1}) experiment at Sp\={p}S, CERN, which started data-taking in 1981.
Concerning data production, UA1 was designed to deliver around \SI{3}{\mega\byte\per\s}, mainly limited by the speed in writing to magnetic tape~\cite{art:astbury}.
The data acquisition of ATLAS on the other hand stores around \SI{320}{\mega\byte\per\s}~\cite{pdf:atlasfact}, with much higher internal data rates.
The increase of a factor 100 in recorded data rate over a time span of 30 years is compensated by the substantial improvement of commercial development in both communication %
and storage. %

Considering the evolution of Ethernet between 1980 and 2010, we have witnessed an increase of about a factor 20 every 10 years in bandwidth~\cite{proc:latha,pdf:ethalliance}, with the major increase in the latter half of the timespan.
After 2010 however, a lower rate of growth, a factor 4 every 10 years, starts to appear.

For data storage, between 1980 and 2010, the increase was on average a factor 30 every 10 years, with a peak between 1990 and 2005 where the area density doubled and prices per byte fell by half on a yearly basis~\cite{art:morris}.
Also this pace has slowed down since around 2010, with instead a factor 4 every 10 years~\cite{art:wood,art:nord}.

This slowdown in industry development poses new data acquisition challenges for both transmission speed and storage.
A particular case when these are in high demand is when scientists are interested in storing entire traces, i.e.\ raw data directly from flash-ADCs, for example during testing or debugging of detectors and data processing procedures.
In this case, the amount of data is much larger, easily by a factor 20--1000 \cite{art:gretina}.

\IEEEpubidadjcol

One way to cope with these challenges is to increase capital expenditure to buy newer and better performing equipment.
However the need to reduce costs leads to a different approach, where we aim to reduce the size of the data to be handled.
This can be achieved through \revlabel{revnoempthdatacompr}data compression.

\revlabel{revtracedescription}
A typical example of the traces considered is time-series data from
flash-ADCs, which usually are slowly varying, with short intervals of
larger variations due to pulses.  The series data can also be information from
adjacent channels, e.g.\ coupled strips of Si detectors,
which can exhibit similar correlation characteristics.

If compression is employed as software running on a PC, only data which has already been sent from the signal acquisition unit can be reduced.
This gives no reduction in the transfer rate demands.
To address both limitations, an implementation of the compression directly on the FPGA, where the initial signal processing takes place, is needed.
This article presents a simple yet effective lossless compression method, that can be applied to sequences of correlated data.
The method allows a straightforward and fast implementation in FPGAs as well as CPUs,
\revlabel{revopensource} and is available as open source software.

This paper is structured in the following way: First, already
available solutions are reviewed, followed by a description of the
present routine.  Optimisation possibilities, both regarding
compression efficiency and resource utilisation are then discussed.
This is followed by descriptions of the interfaces to the FPGA
compression module and the CPU decompression code.  The storage cost
of both noise and pulses are then modeled, and verified using
synthetic trace simulations.  Finally, the achieved storage cost
reduction is benchmarked
using traces from actual detectors.

\section{Overview of available solutions}

Ideas for data compression on front-end electronics are not new.
Scientists working on large detectors have already faced the problem of how to efficiently compress data, albeit with different boundary conditions than in our case.
Both \emph{lossy} compressions, where a part of the initial information is lost to accomplish a reduction~\cite{art:falchieri,art:nicolaucig};
and \emph{lossless} compressions, where the initial information can be fully reconstructed, can be achieved following different approaches.
One is to discard parts of the signal with no or little information (\emph{zero-suppression}~\cite{art:werbrouck}).  %
Another approach is to use a \emph{variable length coding}~\cite{book:khalid}, such as \emph{Huffman coding} \cite{art:huffman} as shown in~\cite{art:mazza} or \emph{Golomb-Rice coding}, which is used in~\cite{art:ammendola}.
The effectiveness of such algorithms is based on the knowledge of the probability distribution of the original data values.
Usually this knowledge is gained from inspecting the whole or a representative pool of the data undergoing compression.
This requires to store and to analyse a representative sample of the data during setup, in order to tune the compression configuration to the signal and ADC operation parameters.
As the signal characteristics have a tendency to change within and between calibration and production data, \revlabel{revopinconvenience}
causing operational inconveniences, such approaches are not suitable for our purpose as a generic configuration-free compression method \revlabel{revgenericfortraces}
for traces, as it causes additional work when operating detectors.

In some cases, through a \emph{pre-processing} of the incoming data, a more advantageous probability distribution can be exploited.
A common approach is the calculation of differences between values~\cite{art:patauner,art:duda,art:kobayashi,art:badier,art:biasizzo}.
These differences may be between sampled data and a model~\cite{art:patauner} or between sampled data and a reference value (base)~\cite{art:duda,art:kobayashi} or between consecutive samples~\cite{art:patauner,art:kobayashi,art:badier}.
When dealing with signal traces, which are sampled at rates high enough that consecutive samples have values close to each other, i.e.\ are correlated, the latter approach delivers a distribution dominated by small values.

\section{Operating principle}

The difference predicted trace compression (DPTC) presented in this paper is based on preserving only those least significant bits which hold the information necessary to recover each value.
Although this does not correspond to a real Huffman coding, the result is to encode the more common smaller values, i.e. closer to zero, with shorter sequences.
This approach is quite similar to the one presented in~\cite{art:kobayashi}, where one sample works as base value and the following three samples undergo the differencing treatment.
The base value can be chosen arbitrarily. We use the first value of each trace, with all following samples subject to the difference processing.
The resulting differences are organised in groups of four, and all the samples in one group are stored using the same number of bits.
A small header containing information about the encoding is placed at the beginning of each group.

\begin{figure}[t]
    \centering
    \includestandalone[width=0.9\linewidth]{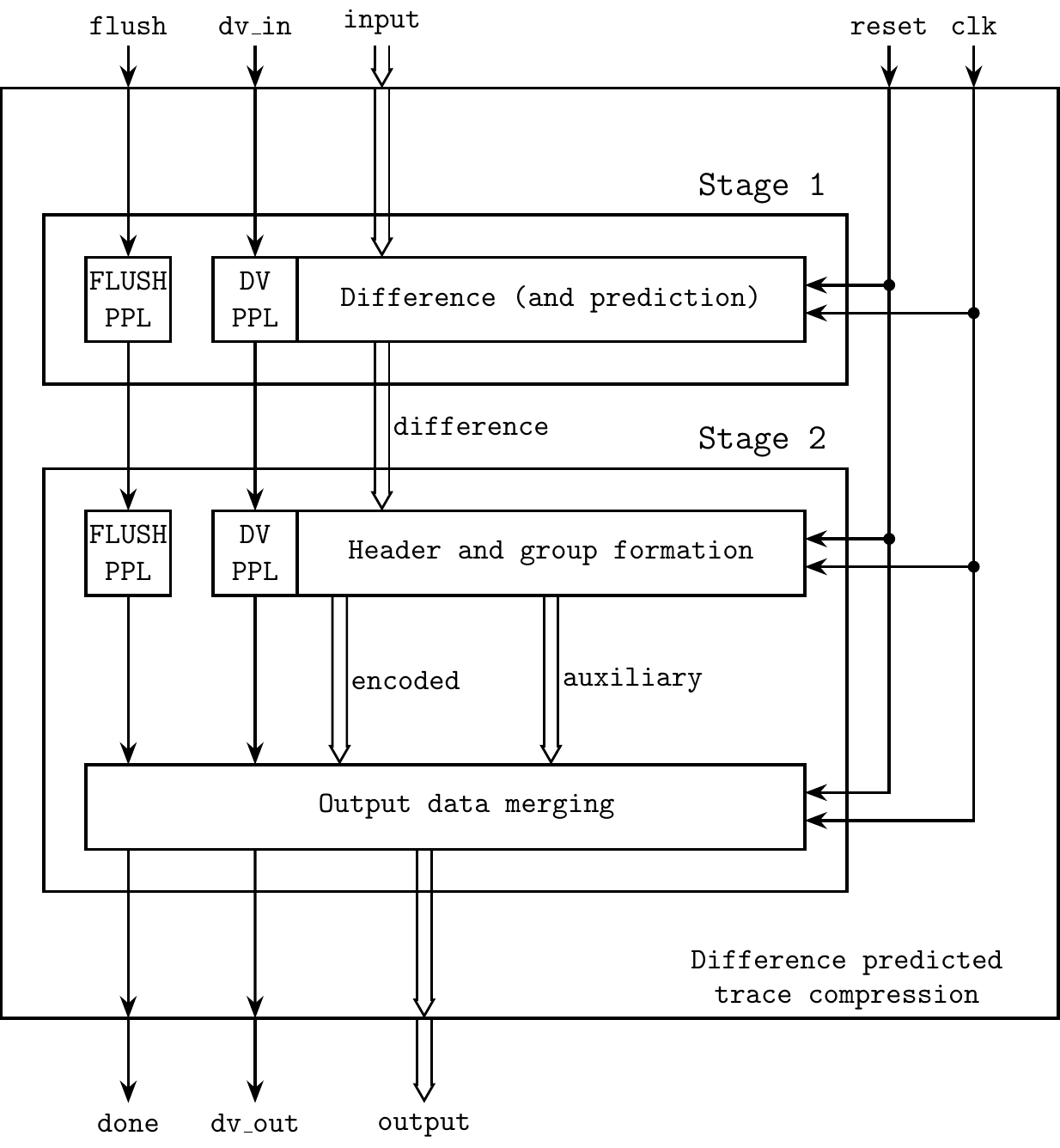}
    \caption{Behavioural structure of the circuit implementing the DPTC algorithm.  The first stage prepares the values to be encoded as differences of the input values.  The second stage concerns the bit-packaging and determines the number of bits needed for each group, and prepares and shifts group headers and encoded values, merging them into the output words.
      One input value can be accepted each cycle, marked by the data valid
      (\texttt{dv}) signal.
      This predicate follows the data pipeline (PPL), and could in the future allow the circuit
      to operate even if new values are not provided every cycle.
      Full \texttt{output} words are signalled by \texttt{dv\_out}.
      The end of a trace is to be marked by the \texttt{flush} signal,
      which, after passing the pipeline,
      ensures that the last data word is generated, followed by
      \texttt{done}.
    }
    \label{fig:module}
\end{figure}

Our implementation is organised in two steps, as shown in \cref{fig:module}.
First, the procedure calculating the differences is applied to the input data.
The original samples consist of a sequence of \emph{n}-bit data words, where \emph{n} is given by the bit resolution of the sampling ADC.
The current design allows \emph{n} to have any value in the range 5--16. %
The second stage is responsible for packing the differences into a stream of 32-bit words.

\subsection{Differencing procedure}
\label{sec:diff}

The first stage treats each value according to the following rules:
\begin{enumerate}
\item Calculate the difference to the previous value.%

\item
  The later storage is due to the binary encoding slightly \emph{asymmetric}. With a
  certain number of bits, it can store one more negative value than positive.
  With e.g.\ 3 bits, the eight differences $-4, -3, -2, \ldots, 3$ can
  be stored.
  For flat (noise-like) parts of a trace, any deviation from zero
  will generally be followed by a difference of the opposite sign.
  To make negative values more common than positive,
  a sign-changing scheme is applied:
  If a stored value is negative, the next non-zero value is stored
  with inverted sign;
  while, if positive, the next is stored as is.
  A value of zero does not change how to store following values.
\end{enumerate}

Note that in all operations, only $n$ bits are considered,
i.e.\ the differences are allowed to wrap (arithmetic is modulo-$2^n$).
This does not introduce any ambiguity.

\subsection{Group creation}
\label{sec:chunk}

\begin{figure}[t]

\centering

{Short encoding, ${\Delta}m=-1, 0,$ or 1.}
\smallskip

\begin{bytefield}{29}
\footnotesize
\bitbox[rlt]{5}{Header} & \bitbox[rlt]{24}{Data values}\\
\scriptsize
\bitbox[lbr]{5}{2 bits} & \bitbox[lbr]{6}{$m$ bits} & \bitbox[lbr]{6}{$m$ bits} & \bitbox[lbr]{6}{$m$ bits} & \bitbox[lbr]{6}{$m$ bits}\\
\scriptsize
\bitbox[lrb]{5}{01, 10, 11} & \bitbox[lrb]{6}{1st diff.} & \bitbox[lrb]{6}{2nd diff.} & \bitbox[lrb]{6}{3rd diff.} & \bitbox[lrb]{6}{4th diff.}\\
\end{bytefield}

\bigskip

{Long encoding, other ${\Delta}m$.}
\smallskip

\begin{bytefield}{35}
\footnotesize
\bitbox[rtl]{11}{Header} & \bitbox[rlt]{24}{Data values}\\
\scriptsize
\bitbox[lbr]{5}{2 bits} & \bitbox[lbr]{6}{$k$ bits} & \bitbox[lbr]{6}{$m$ bits} & \bitbox[lbr]{6}{$m$ bits} & \bitbox[lbr]{6}{$m$ bits} & \bitbox[lbr]{6}{$m$ bits}\\
\scriptsize
\bitbox[lrb]{5}{00} & \bitbox[lrb]{6}{${\Delta}m-2$} & \bitbox[lrb]{6}{1st diff.} & \bitbox[lrb]{6}{2nd diff.} & \bitbox[lrb]{6}{3rd diff.} & \bitbox[lrb]{6}{4th diff.}\\
\end{bytefield}

\caption{Layout of a group of difference values together with its header.
  Depending on the difference ${\Delta}m$ in the number of bits that are needed in this group compared to the previous, the group header is characterised by a \emph{short} (upper) or \emph{long} (lower) encoding. The parameter $k$ is fixed by the maximum number of bits that are needed to represent the maximum difference not covered by the short header $\max({\Delta}m) = n-3$. %
}
\label{fig:bitschunk}

\end{figure}

\begin{figure}[t]
    \centering
    \includegraphics[width=\linewidth]{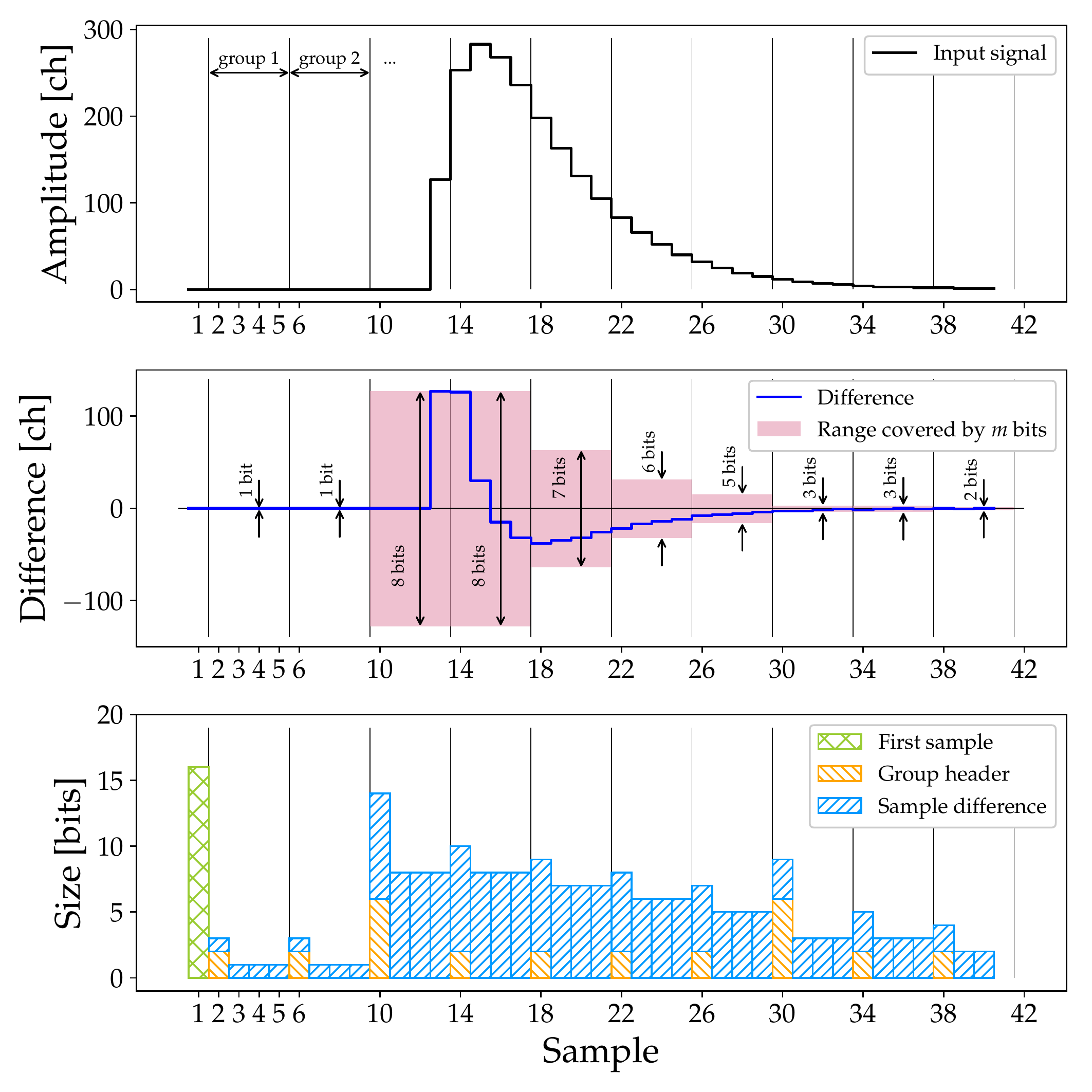}
    \caption{
      Graphical representation of the differencing and group
      formation procedures.
      Shown in the upper panel is a simulated double-exponential signal
      (difference of double exponentials, DEXP~\cite{art:wu}),
      which is used as input data, with $n=16$.
      The center panel shows the differences, and how the number of
      bits, $m$, used in each group, depends on the largest difference.
      The lower panel shows the compressed data size in bits.
    }
    \label{fig:single_cnk}
\end{figure}

The values are stored in groups of four, using the same number of bits, $m$, for each value in a group.
This is illustrated in \cref{fig:bitschunk,fig:single_cnk}.
Since the stored values are differences, both positive and negative values must be representable (in \emph{two's complement} representation).
Since each value may require a different number of bits to be represented,
the widest representation needed by any value in a group is used.
The number of bits used for values in each group is stored in a group header, placed before the actual data.
Considering consecutive groups, it is worth noticing that the number of bits needed will often not change much and therefore a \emph{short} and \emph{long} encoding of the number of bits is employed, see \cref{fig:bitschunk}.
The short header consists of two bits: if the encoded value is 1, 2, or 3, the number of bits to use for the group is the same as for the previous group with a change of $-1$, $0$ or $+1$ bits, respectively.
If the value of the two-bit short header is 0, the encoding is long and contains the full difference of bits stored per value.
Since some values are already covered by the short encoding, an offset of 2 is applied to the full difference.
This is encoded using $k$ bits, which is chosen such that any needed difference, at most $n-3$, can be stored;
\revlabel{revformulak}%
$k = \lceil \log_2 (n-3) \rceil$,
i.e. 1 bit for $n = 5$, 2 bits for $n \leq 7$, 3 bits for $n \leq 11$, and 4 bits for $n \leq 19$.

The number of bits per stored difference is interpreted with a bias of 1, meaning that storing 0 bits per value is not supported.
This is a conscious choice: supporting 0-bit values (i.e.\ minimal encoding
of groups
with all value differences 0) would make the code for CPU decompression
(and compression) more complicated.
Since ADCs usually are operated with noise in the least significant bit,
it is also expected to have limited practical use.

The data values are then stored with the necessary number of bits for the group.
Each data value is stored with a bias relative to the most negative value that can be stored with the given number of bits.
This simplifies decoding, as the stored value only has to be unmasked, and the bias subtracted.
This avoids a cumbersome sign extension operation by the CPU decoder.

As an exception to the above rules, the first data value is stored alone and fully, using \emph{n} bits.
This avoids storing the entire first group of data with many bits.

\subsection{Output word formation}
\label{sec:shifting}

The resulting stream of bits is then packed in 32-bit words, being filled from the least significant bits.
When a value to store cannot fit, the completed output word is emitted and the remaining bits are stored in the next 32-bit output word.

Information about the number of original data values, number of data words produced by the compression and $n$ is needed by the decompression procedure.
These values are not recorded by our routine, therefore it is the responsibility of the user to retain this information.

\section{Optimisation}

The algorithm described in the previous section can be optimised in different ways.
However, the improvements obtained by applying additional procedures depend on many aspects, such as noise level, signal shape, and the distribution of signal amplitudes.
Note that while improving for some characteristic, an optimisation will undermine other aspects.
We present a few ideas together with a short analysis of each one, discussing advantages and disadvantages.

\subsection{Compression factor optimisation}

\subsubsection{Linear predictor}
With this additional pre-processing, the linear component of long sloped parts of a trace are removed by a second differencing of the data.
This aims at a distribution of values more narrow %
around zero.
However, for flat parts of a trace, which mainly contain noise, such a double difference leads to a wider distribution.
Thus, in order to give an overall improvement, this procedure must only be applied for sufficiently long, sloped sequences.
This is controlled by a heuristic using the observation that consecutive samples in unfavorable regions change sign often, or have 0 difference, and thus can be detected by a three-most-recent rule.
The second differencing is switched off when at least one sign change or a zero has occurred for the previous three values.

While at first appearing to be promising for synthetic traces,
from tests on actual traces, this optimisation does however not bring any improvement.
This is connected to the fact that usually most of the pulses in the digitised traces only have small amplitudes, therefore the few improvements by this predictor are neutralised due to it activating spuriously in flat parts.
The optimisation is implemented in the code, but deactivated by default.

\subsubsection{Number of values in a group}

\begin{figure}[t]
    \centering
    \includegraphics[width=0.95\linewidth]{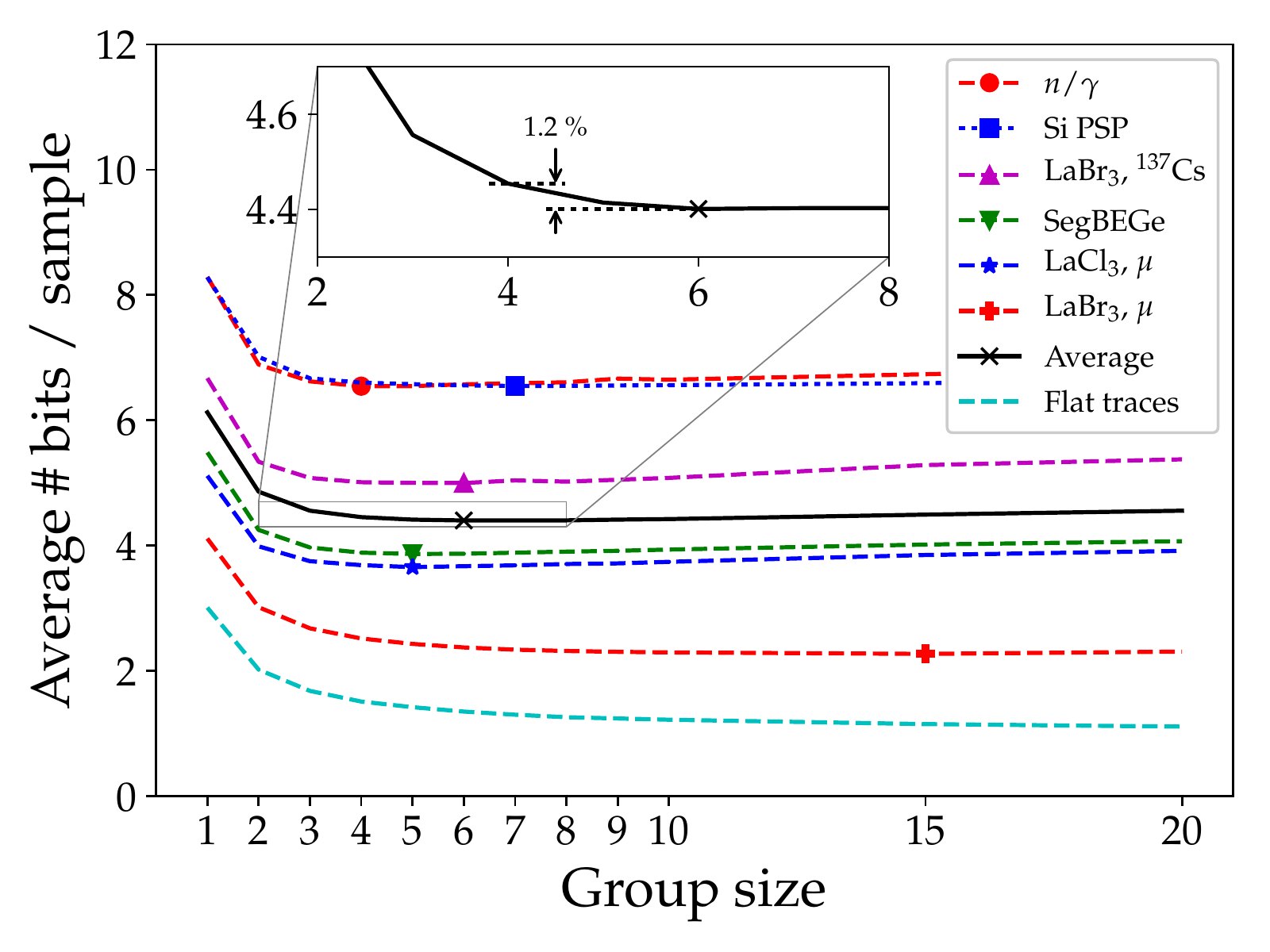}
    \caption{%
      Compression efficiency as a function of the group size for the
      actual traces presented later in \cref{tab:realtraces}.
      The results have been averaged (geometrically) within the different
      sample groups, and then again to provide a total average.
      The minimum locations are indicated.
      The flat test traces are not included in the total average.
    }
    \label{fig:chunksize}
\end{figure}

The group size can also be varied to optimise the compression
efficiency, see \cref{fig:chunksize}.  Using smaller groups require more storage space
due to the more frequent headers, while larger groups will encode
unnecessary bits for more samples.  The figure shows an optimum
around six samples per group, with gradual losses at larger values, or
steeply below three.

We have chosen to code four values in each group.  The loss is about
\SI{1.2}{\percent} compared to groups of six values.
Fixing the number as a power of two might be useful
for a future parallelized unpack code.
\revlabel{revchoosefour}%
We choose four rather than eight as this leads to shorter
pipelining in the group formation part of the circuit.

\subsection{Circuit optimisation}

\subsubsection{Additional pipeline stages}
\label{sec:pipeline}
The achievable minimum clock cycle period in a digital circuit depends on the
propagation delay of the longest combinational logic chain between register latches.

In our case the circuit is described in VHDL, where the model
and grade of the FPGA that is targeted will affect which logic expression becomes the longest.
Adding pipeline stages to split the longest paths helps to lower the minimum clock cycle.
At the same time however, introducing a pipeline stage causes more LUTs\footnote{Look-up table, a basic FPGA building block.  \revlabel{revlutnotff}The other basic unit is signal registers, i.e.\ flip-flops (FF).} to be used, as well as flip-flops; leading to a trade-off between resource-usage and speed.
In order to allow flexibility when using the code, a few generic parameters control a number of optional pipeline stages. %

Since the synthesized code uses more LUTs than flip-flops,
compared to the usually available ratio on FPGAs, we concentrate on
the LUT usage for the circuit optimization comparisons.

\begin{figure}[t]
    \centering
    \includegraphics[width=\linewidth]{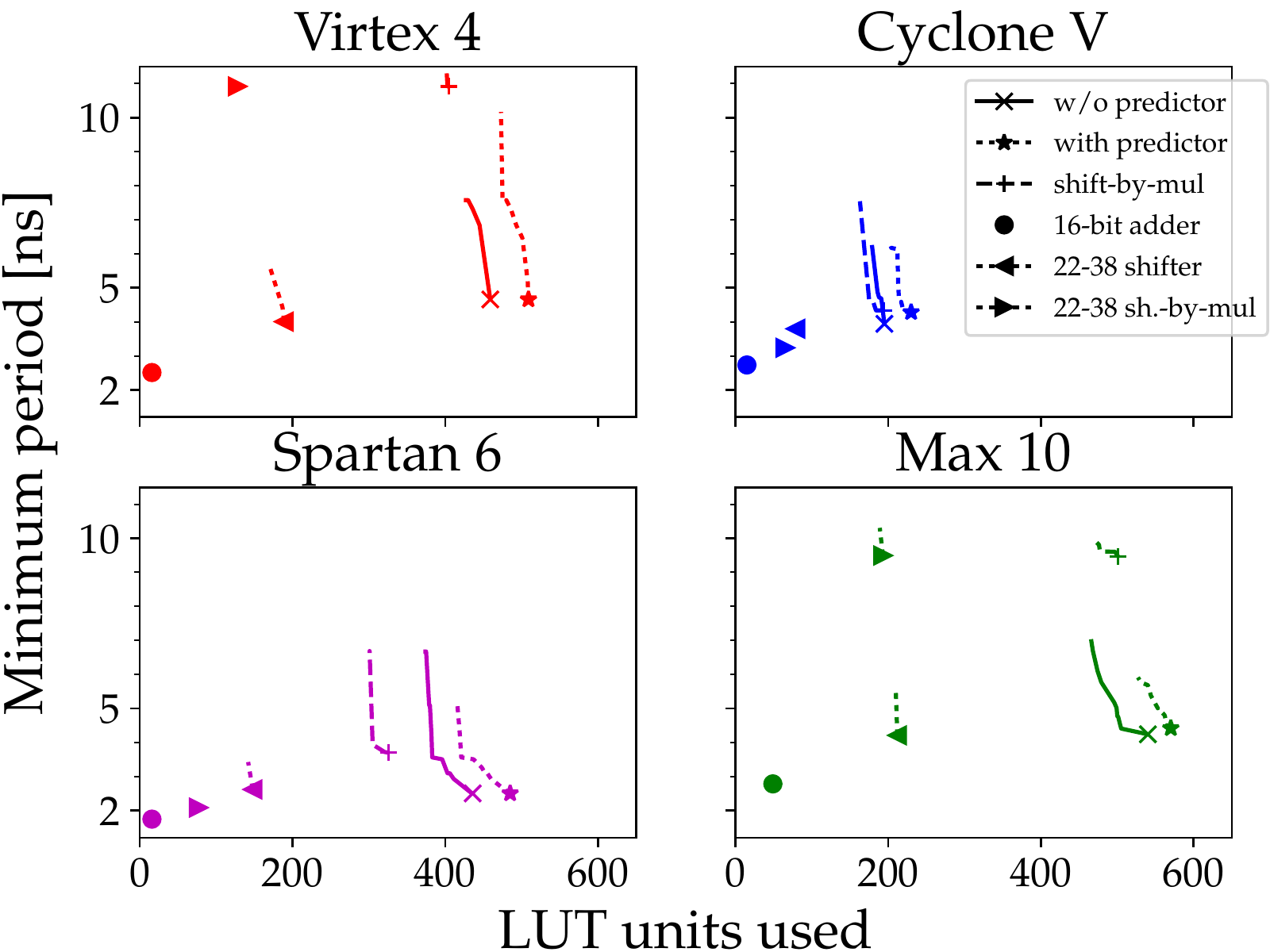}
    \caption{Achievable circuit frequency as a function of look-up table (LUT) usage
      for some commercial FPGA architectures
      from the companies Xilinx (left column) and Altera (now Intel; right column),
      mainly varied by additional pipeline stages.
      The circuit handles one data values per clock cycle,
      indendent of the actual value.
      Three main configurations, all with $n=16$, are evaluated:
      without and with the predictor, and by doing the shift using
      multiplier units.
      The resource use of a single 16-bit adder, and a 22-bit 38-position shifter are also shown.
      Note that different FPGA models have resources (e.g.\ LUTs) with different capabilities, thus the resource consumption cannot be meaningfully compared between models.
    }
    \label{fig:freq_vs_luts}
\end{figure}

By performing VHDL synthesis for all combinations of the optional pipeline
stages, and directing the respective FPGA development toolchain to optimise for speed, the
achievable performance as function of resource usage can be determined.
The results are shown in \cref{fig:freq_vs_luts} and summarized in \cref{tab:clockocc}.
\revlabel{revdescribefreqvslutsfig}%
Locations further down in the figure indicate that shorter clock
periods can be used, and further to the left mean less resource consumption.
For each circuit, only the results which improve the achievable
clock frequency for a certain resource usage is kept,
thus the short curves mainly show the improvements possible as more
pipeline stages are enabled.  To a smaller degree they also come from
the ability of the toolchains to trade resource usage for speed.
\revlabel{revmostused16add}%
To compare with the most used constructions (adders, subtractors,
comparators), 16-bit adders are also shown in the figure.
The VHDL code
allows the minimum period of the clock to be below
\SI{10}{\ns} (i.e. \SI{100}{\MHz}) even on 10-year old FPGAs,
and it can easily be configured to reach below \SI{5}{\ns} with additional pipeline stages.
On more modern FPGAs, going below \SI{3}{\ns} seems rather easy.
\revlabel{revfastasfadc}
If the compression circuit is operated continuously, directly fed by
the data generator (e.g.\ flash-ADC), the speed needs to match the
sampling period, since the circuit can process one sample per clock
cycle.
When compressing only selected traces which
first have been recorded into temporary memory buffers,
a slower clock can be used for the compression circuit.

\begin{table}[t]
  {
    \centering
    \caption{Circuit resource usage for different FPGA targets.
    }
    \label{tab:clockocc}
    \begin{tabular}{lccc}
      \toprule
      \textbf{FPGA model} & \textbf{LUT} & \textbf{FF} & \textbf{Max clock frequency}  \\
                   & \textbf{occupation} & \textbf{occupation} & {\si{MHz}}  \\
      \midrule
      Xilinx Virtex 4	& 427--459	 & 265--301 & 210 \\ %
      \medskip
      Xilinx Spartan 6	& 374--436	 & 234--300 & 400 \\
      Altera Cyclone V	& 179--195	 & 262--324 & 250 \\ %
      Altera Max 10	& 466--540	 & 220--308 & 230 \\
      \bottomrule
    \end{tabular}\par
  \bigskip
  }
  \cref{tab:clockocc}
  shows the resource usage for configurations with $n=16$, without the predictor.
  Note that different FPGA models have resources (e.g.\ look-up tables (LUT)) with different capabilities, thus the resource consumption cannot be meaningfully compared between models.  The number of signal registers, flip-flops (FF), is also given.  This includes $19+34$ FFs as inputs to, and outputs from, the actual module.
\end{table}

The single most expensive component of the circuit is the barrel
shifter, which aligns the encoded data at the next position in the output word.
For $n=16$, the shifter input is 22 bits wide, with the additional
6 bits coming from the potentially long encoded header.
The shift amount is in the range 0 to 37, inclusive.
0 to 31 depends on how many bits already are used in the output word.
The additional 0, 4, or 6 positions depend on the header
(long, short, or none). %
This gives a 60 bit output.
\revlabel{revmentionbareshifterfig}%
The cost and performance of the shifter units are also shown in
\cref{fig:freq_vs_luts}.

\revlabel{revmovesectIVb2}%
\subsubsection{Barrel shifter vs. multiplier units}

A barrel shifter on FPGAs is normally realised as one multiplexer
for each output bit (sharing some parts of the first stage selectors of each multiplexer).
Since it also can be expressed as a multiplication of the input value with $2^i$,
where $i$ is the shift value,
it can also be implemented using multiplier units in
FPGAs. For the second factor, the input value is generated as $2^i$,
i.e.\ a one-hot encoding of the shift amount.

One could imagine this to be beneficial when generic LUT resources are scarce, however for the cases tested,
it is not.  The generation of the $2^i$ input value is rather
expensive, as it requires $2^i$ individual selectors.  Also the combination of
the output values from the several multiplier units, often $9 \times 9$ or $18 \times 18$ wide, are rather expensive.

The resource usage for 22-bit, 38-position left-shifters implemented
in the two ways are also compared in \cref{fig:freq_vs_luts}.

\revlabel{revlutvsalln}
The results in \cref{fig:freq_vs_luts} and \cref{tab:clockocc} are for
$n=16$.  Similar tests for $5 \le n < 16$ give that for each bit removed,
the needed number of LUTs shrinks on average by \SIrange{4}{5}{\percent},
and the attainable minimum period required decreases by
\SIrange{1}{2}{\percent}, depending on FPGA model.

\section{VHDL module interface} %

The interface to the VHDL compression module is a single entity, with
input and output signals as seen at the top and bottom of \cref{fig:module}.
Optional pipeline stages are configured using a generic map.

The circuit inputs are:
\begin{itemize}
\item \texttt{clk}:
  clock signal;
  
\item \texttt{reset}:
  reset signal,
  given for at least as many cycles as the pipeline has stages;

\item \texttt{input}:
  $n$-bit data value to compress;
  
\item \texttt{dv\_in}:
  data valid signal: set to '1' every clock cycle an input value is provided;
  
\item \texttt{flush}:
  flush signal: set to '1', after the last input value has been
  given.  Held until \texttt{done} reported back.
  This forces the last output word to be emitted, especially when it
  is not fully occupied.
\end{itemize}

The output signals are:
\begin{itemize}
\item \texttt{output}:
  32-bit output data word;
  
\item \texttt{dv\_out}:
  data valid signal: '1' every time the output word is filled,
  signaling the presence of a completed data word to be stored;
  
\item \texttt{done}:
  informs that the last input value has been processed and the
  final output word was produced (possibly in the current cycle).
\end{itemize}

\section{Decompression}

The decompression is performed by one C function with the following parameters:
\begin{itemize}
\item 
\texttt{compr}:		pointer to the 32-bit words of the compressed input buffer;
\item
\texttt{ncompr}:	number of elements in the input buffer;
\item
\texttt{output}:	pointer to a buffer of 16-bit items for the decompressed values;
\item
\texttt{ndata}:		number of original/decompressed values;
\item
\texttt{bits}:		number of bits of each value that was stored ($n$).
This must be the same as the number configured during compression.
\end{itemize}
On success, 0 is returned, otherwise a non-zero value.

\begin{table}[t]
  {
    \centering
    \caption{Performance of the decompression routine on various CPUs.}
    \label{tab:decompression}
    \begin{tabular}{lS[table-number-alignment = center]cS[table-number-alignment = center]@{\hskip 0mm}c@{\hskip 0mm}S[table-number-alignment = center]}%
      \toprule
      \textbf{CPU Model}  & \textbf{Speed} & \textbf{Released} & \multicolumn{3}{c}{\textbf{Time/sample}}\\
                          & {(\si{GHz})} & & \multicolumn{3}{c}{{(\si{ns})}}\\
      \midrule
      Xeon E3-1285v6      & 4.5 & 2017 &  2.0 &--&  2.2 \\
      Xeon E3-1276v3      & 4.0 & 2014 &  2.5 &--&  3.1 \\
      Xeon X5450          & 3.0 & 2007 &  5.3 &--&  6.2 \\
      PPC 7455            & 1.0 & 2002 & 28   &--& 36   \\
      \bottomrule
    \end{tabular}\par
    \bigskip
  }
  \cref{tab:decompression} shows the single-threaded decompression times per sample for
  the actual traces presented later in \cref{tab:realtraces}.
  With 16-bit data samples, and a decompression time on modern
  hardware smaller than \SI{2.5}{\ns\per sample}, the decompressed
  rate is larger than \SI{800}{\mega\byte\per\s}, i.e.\ well
  comparable to solid state drives (SSDs).
\end{table}

The routine will report decompression failure on malformed compressed data,
e.g.\ if there are non-zero bits left in the input buffer, or when entire
words have not been used.
The decompression routine will not read items beyond the end of the
source buffer even if it runs out of data, e.g.\ due to a corrupted
data stream.
\Cref{tab:decompression} shows the typical performance,
which
only has a small dependence on the actual data values.

\section{Compression efficiency---Storage cost}

The contributions to the compressed data size can be divided in two
parts:

\begin{enumerate}
\item
  The cost of storing traces with no pulses, i.e.\ only containing the
  digitization noise.
  This is described as a cost per sample.

\item
  The cost of storing a pulse,
  described as an additional cost for the entire pulse.

\end{enumerate}

There is a natural interplay between the two, as the noise affects the
additional cost to store a pulse.  This effect is also addressed below.

In the following, we use the variables $c$ for cost and $b$ for bits.
To specify these, subscripts are used:
$\mathcal{N}$ for noise,
$\mathcal{T}$ for trace,
$\mathcal{S}$ for sample,
$\mathcal{P}$ for pulse, and
$\mathcal{B}$ for a small pulse (bump).
Gaussian noise is described by its amplitude $\sigma_{\mathcal{N}}$.
The amplitude and width (std. dev.) of Gaussian-shaped pulses are given by $A_{\mathcal{P}}$ and $w_{\mathcal{P}}$.

\subsection{Bare trace cost}
\label{sec:tracecost}

The cost of storing a trace without pulses has two parts: the
size of the headers and the size of the encoded values, i.e.\ the differences.

The cost of storing the differences depends on the noise content, most
easily expressed as the number of bits of noise
$b_\mathcal{N} = \log_2 \sigma_{\mathcal{N}}$.

\begin{figure}[t]
  \centering %
  \includegraphics[width=0.95\linewidth]{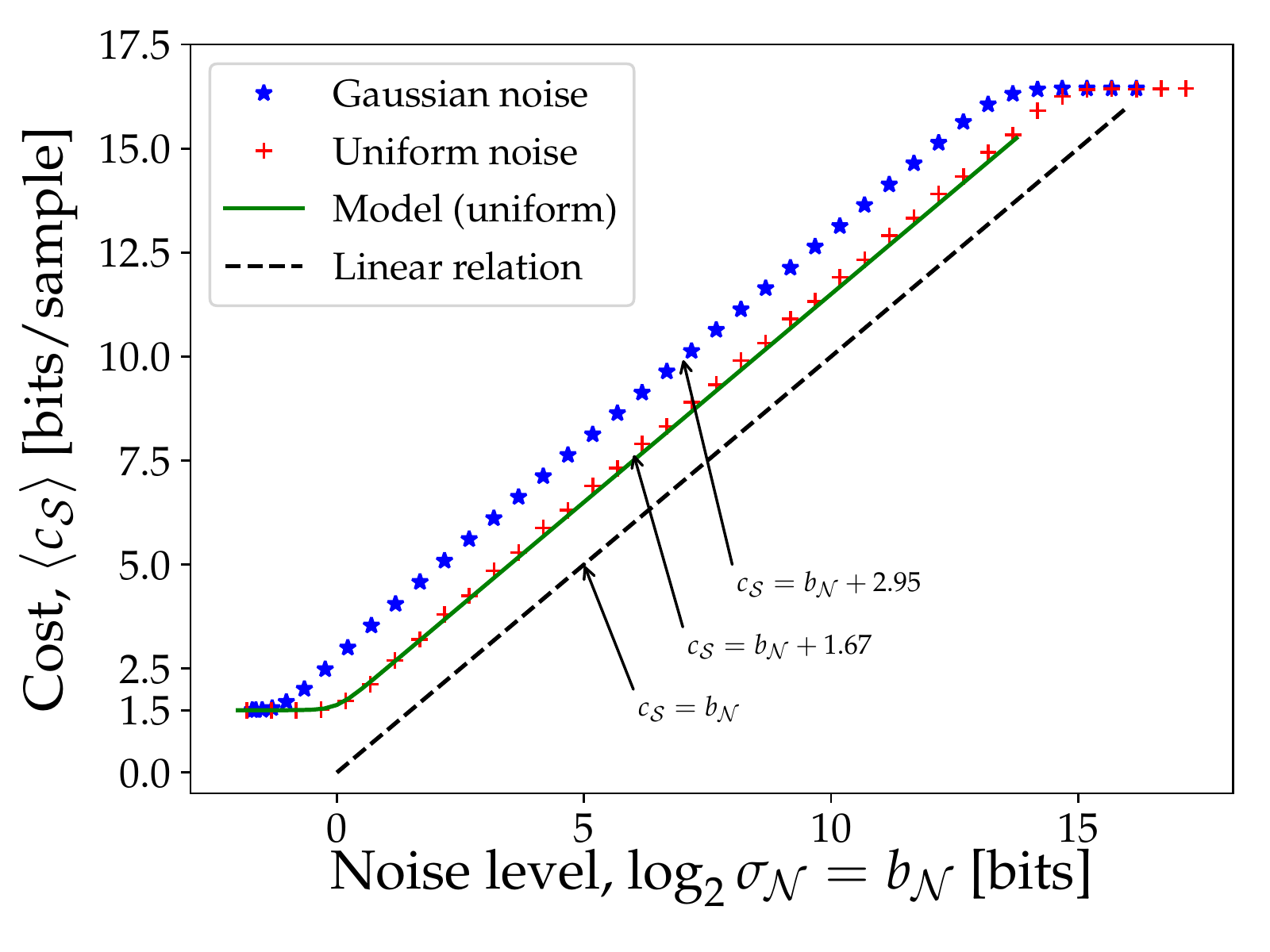}
  \caption{Average cost per sample depending on the number of noisy bits.
    The minimum cost per sample is 1.5 bits, given by one bit per sample and a short header of two bits every group of four samples.
    The model is compared to actual compression of traces with
    either Gaussian or uniform noise.
    For Gaussian noise the standard deviation of the distribution expresses the number of noisy bits.
    For uniform noise instead the number of noisy bits represents the span of the random value distribution.
    }
  \label{fig:comprsample}

\end{figure}

Ignoring the pecularities of the first group,
which may require a long header encoding, the estimated cost for a trace
$\avgbr{c_\mathcal{T}}$ will be proportional to its length $n_\mathcal{T}$:
\begin{linenomath*}
  \begin{equation}
    \avgbr{c_\mathcal{T}} = n + (n_\mathcal{T}-1) \avgbr{c_\mathcal{S}} + 15.5.
  \end{equation}
\end{linenomath*}
The first sample has a fixed cost $n$.
The constant 15.5 accounts for the average number of unused bits
in the last output word at the end of a trace.
A first approximation, denoted by the tilde, for
the average cost per noise sample is
\begin{linenomath*}
  \begin{equation}
    \avgbr{\widetilde{c_\mathcal{S}}} = 0.5 + b_\mathcal{N} + 1 + o.
  \end{equation}
\end{linenomath*}
The first half bit comes from the short group header, using two bits every four
samples.
The additional one comes from differences encoding both positive and negative entries, i.e.\ effectively a sign bit. %
The term $o$ is an overhead, since the grouping of values causes some more bits
than necessary to be used. %
\revlabel{revtotalcostlownoise}%
To model the transition from very small noise levels,
where the total cost is \SI{1.5}{bits/sample},
to the proportional regime,
\revlabel{revsmoothtransition}%
a smooth transition function $g(x) = \frac{1}{f}\log_2 (1+x^f)$
is used for $b_\mathcal{N}$, with $x = \sigma_{\mathcal{N}}$.
As wanted, $g(x) \to 0$ as $x\to{0}$ and
$g(x) \to \log_2 x$ for $x\gg{1}$, while
the parameter $f$ controls the smoothing.
This yields:
\begin{linenomath*}
  \begin{equation}
    \avgbr{c_\mathcal{S}} = 0.5 + \frac{1}{f}\log_2 (1+{\sigma_{\mathcal{N}}}^f) + 1 + o.
  \end{equation}
\end{linenomath*}
This is illustrated for Gaussian and uniform noise in \cref{fig:comprsample},
\revlabel{revcontrolparamf8}%
where good fits are achieved with $f=8$.
\revlabel{revmovefig6caption}
    For uniform noise, the range of differences is twice as large as the value distribution (due to also encoding negative entries), explaining the use of, on average, one more bit per sample in addition to the short header and $b_\mathcal{N}$.
    This is modeled by (3) shown as a solid line.
    For Gaussian noise, the distribution of differences between
    consecutive samples is wider by a factor $\sim\sqrt{2}$ than the
    original distribition, and any large value in a group of four leads
    to longer encodings.
    The further small fractional costs per sample are in both cases
    likely given by the occasional use of long group headers.

\subsection{Pulse cost}
\label{sec:pulsecost}

The cost of a pulse is best described as the total cost of the pulse,
and not a cost in bits per sample.  For the following discussion, 
pulses are assumed to have a Gaussian shape, \revlabel{revgaussshape}
as opposed to the double exponential function considered in \cref{fig:single_cnk}.
Detector pulses can be considered as composed of two parts with
different time constants (i.e.\ widths)
for the rising and falling parts.
Even if this may be a rather
rough approximation of real pulses, especially for the trailing part, 
it is practical, \revlabel{revgaussshapetwo}
since Gaussian functions are efficiently and familiarly described using their widths and amplitudes.

Since it is differences that are stored, the important parameter is
not the amplitude $A_\mathcal{P}$ of a pulse, but its steepest
slope, which scales as $\frac{A_\mathcal{P}} {w_\mathcal{P}}$.  As
a first approximation, denoted by the tilde,
the cost is proportional to the number of
bits needed to store these differences, as well as the width of the pulse,
\begin{linenomath*}
  \begin{equation}
    \label{eq:pulsecostbadlimitbehaviour}
    \avgbr{\widetilde{c_\mathcal{P}}} =
    a w_\mathcal{P}
    \log_2 \left( \frac{A_\mathcal{P}} {w_\mathcal{P}} \right).
  \end{equation}
\end{linenomath*}
The scale is given by the proportionality constant $a$.
It turns out that this formula works rather well, if modified to account for
the facts that even for small pulses, costs are not negative (by adding 1 inside the logarithm and control parameter $b$),
and that very narrow pulses still will affect the storage size of
at least one entire group ($d$ within the square root):
\begin{linenomath*}
  \begin{equation}
    \label{eq:pulsecost}
    \avgbr{c_\mathcal{P}} =
    a \sqrt{w_\mathcal{P}^2 + d^2}
    \;\frac{1}{b}
    \log_2 \left( 1 + \left( \frac{A_\mathcal{P}} {w_\mathcal{P}} \right)^b \right).
  \end{equation}
\end{linenomath*}
The modification is thus adjusted by the control parameters $b$ and $d$.

\subsection{Pulse-noise interaction}
\label{sec:pulseinteract}

The above description \cref{eq:pulsecost} works in the limit where the pulse is
large compared to the background noise.  When this is not the
case, the additional cost of storing the pulse will be \emph{smaller}, since
the pulse-associated part of the differences to some extent will be
covered by the noise storage cost.
This can be modeled by
\begin{linenomath*}
  \begin{equation}
    \label{eq:smallpulsecost}
    \avgbr{c_\mathcal{B}} = \sqrt{\avgbr{c_\mathcal{P}}^2 + \avgbr{c_\mathcal{NB}}^2} - \avgbr{c_\mathcal{NB}}.
  \end{equation}
\end{linenomath*}
The correction is the cost of storing the noise for a stretch of
samples proportional to the pulse width:
\begin{linenomath*}
  \begin{equation}
    \avgbr{c_\mathcal{NB}} = q \sqrt{w_\mathcal{P}^2 + d^2}\; \frac{1}{f} \log_2 \left( 1 + {\sigma_\mathcal{N}}^f \right) .
  \end{equation}
\end{linenomath*}
$q$ is a proportionality constant.

\begin{figure}[t]
  \centering
  \includegraphics[width=0.95\linewidth]{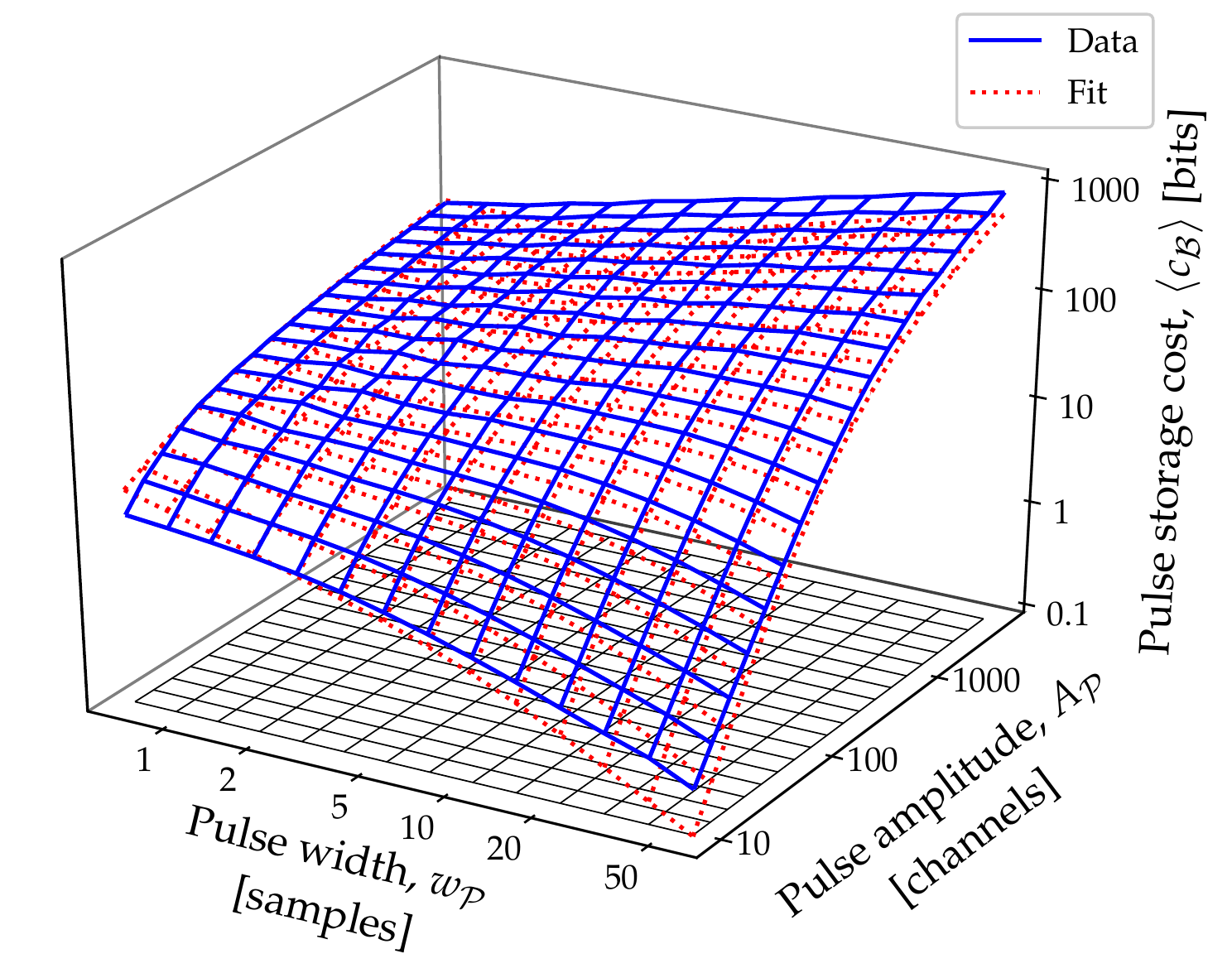}
  \caption{Cost to store a Gaussian pulse as a function of the pulse amplitude and width. This is shown for both compressed simulated data (solid lines) and the model~\cref{eq:smallpulsecost} (dotted lines). The input traces are built adding Gaussian noise, in this case with a sigma of 2.0, to the Gaussian pulse.
    The cost of storing only the Gaussian noise is then subtracted from the cost of storing the trace, leaving the cost of storing the pulse. 
    The fit is done using model~\cref{eq:smallpulsecost} on the total amount of data and minimizing relative differences.
  }
  \label{fig:compr_pulses}
\end{figure}

\subsection{Storage cost verification---synthetic traces}

The storage cost described above and culminating in \cref{eq:smallpulsecost}
has been verified by simulating a
large number of traces with Gaussian pulses, where the parameters
$A_\mathcal{P}$, $w_\mathcal{P}$, and $\sigma_\mathcal{N}$ were
varied.
A global fit suggests the following values for the control parameters:
$a = \num{5.6}$, $b = \num{1.3}$, $q = \num{33}$, $d = \num{2.6}$ and $f = \num{7.7}$,
with a parameter uncertainty of up to \SI{10}{\percent}.
\Cref{fig:compr_pulses} shows the $\sigma_{\mathcal{N}} = 2.0$ case. %

Simulations were performed by building, for each set of parameters, a set of \num{15E6} traces, each made of 500 samples, with Gaussian noise $\sigma_\mathcal{N}$. %
In each, a Gaussian pulse ($A_\mathcal{P}$, $w_\mathcal{P}$)
was added to the trace. %
To average over discretisation effects, both the (noise) baseline and
the center of the pulse were randomised, trace by trace, with fractional offsets.

Although \cref{fig:compr_pulses} shows a good agreement between Equation~\cref{eq:smallpulsecost} and the data, larger differences emerge for small values of $A_\mathcal{P}$ and $w_\mathcal{P}$.
These correspond to the limits handled by the modifications between \cref{eq:pulsecostbadlimitbehaviour} and \cref{eq:pulsecost}, which are thus seen to only partly address these edge effects.

\begin{table*}[t]
  {
    \centering
    \caption{Compression efficiency of the DPTC algorithm for actual traces, categorised by detector type and detected radiation, and compared to popular general-purpose compression methods and Huffman encoding.}
    \label{tab:realtraces}
    \begin{tabular}{lllrrrrrcrcc}%
\toprule
      \textbf{Label} & \textbf{Category}   & \textbf{Details} & \textbf{Traces} & \textbf{Samples} & $\avgbr{A_\mathcal{P}}_\mathrm{g.}$ & $\sigma_\mathcal{N}$ & $\avgbr{c_\mathcal{S}}$ & \textbf{DPTC} & \textbf{gzip} & \textbf{xz(LZMA)}  & \textbf{Huff.} \\
       & \textbf{}           & \textbf{} &         \#              & \#               &                                    &                    &                        & \multicolumn{4}{c}{--- --- --- --- Bits/sample --- --- --- ---} \\
\midrule      
      a&                          & core signal      &  40 &  5000 &  78.3 &  2.16 & 4.06 & \textbf{3.89} &  5.54 &  4.04 & 3.58 \\ %
      b& $\gamma$ in segmented BEGe & segment 1        &  40 &  5000 &  27.3 &  2.16 & 4.06 & \textbf{3.86} &  4.87 &  3.89 & 3.55 \\ %
      c&                          & segment 5        &  40 &  5000 &  53.2 &  2.21 & 4.09 & \textbf{3.91} &  5.54 &  4.13 & 3.59 \\ %
\midrule
      d& \multirow{3}{*}{\begin{tabular}{@{}l@{}}n/$\gamma$ discrimination\end{tabular}}
                        & Ionisation chamber     & 200 &   200 &  907 &  71.2  & 9.10 & \textbf{9.16} & 11.37 &  9.78 & 8.75 \\ %
      e&                & $n$-det. anode         & 200 &   200 &  226 &   4.88 & 5.24 & \textbf{5.36} &  6.63 &  5.32 & 5.12 \\ %
      f&                & $n$-det. cathode       & 200 &   200 &  220 &   6.20 & 5.58 & \textbf{5.71} &  7.00 &  5.62 & 5.46 \\ %
\midrule
      g& position-sensitive & $\alpha$-particles &  50 &  1000 &  852 &   29.7  & 7.84 & \textbf{7.81} & 11.07 &  8.10 & 7.38 \\ %
      h& Si pin-diode       & $^{40}$Ar          &  50 &  1000 &  638 &  6.36 & 5.62 & \textbf{5.58} &  9.37 &  6.23 & 5.24 \\ %
\midrule
      i&                  & no signal split  & 100 &   200 &  534 &  5.30 & 5.36 & \textbf{5.55} &  7.91 &  6.33 & 5.47 \\ %
      j&  $\gamma$ from $^{137}$Cs                                                                                 
                          & signal split 1:2 & 100 &   200 &  292 &  3.90 & 4.91 & \textbf{5.08} &  7.18 &  5.67 & 4.98 \\ %
      k&  in LaBr$_3$     & signal split 1:4 & 100 &   200 &  194 &  3.23 & 4.64 & \textbf{4.81} &  6.69 &  5.24 & 4.68 \\ %
      l&                  & signal split 1:8 & 100 &   200 &  122 &  3.05 & 4.56 & \textbf{4.65} &  6.37 &  5.06 & 4.43 \\ %
\midrule
      m& \multirow{3}{*}{\begin{tabular}{@{}l@{}}cosmic $\mu$ in LaBr$_3$,\\varying HV of PMT\end{tabular}}
                       & 350V $^\mathit{a}$ & 100 &   600 &   9.2 & 0.25 & 0.94 & \textbf{1.67} &  0.65 &  0.49 & 0.65 \\ %
      n&               & 400V $^\mathit{a}$ & 100 &   600 &  19.4 & 0.25 & 0.94 & \textbf{1.67} &  0.84 &  0.63 & 0.78 \\ %
      o&               & 450V               & 100 &   200 &  921  & 4.28 & 5.05 & \textbf{5.55} &  8.36 &  6.42 & 5.76 \\ %
\midrule
      p& \multirow{2}{*}{\begin{tabular}{@{}l@{}}cosmic $\mu$ in LaCl$_3$,\\different digitizers\end{tabular}}
                       & CAEN DT5730 & 100 &   400 &  301  & 3.88 & 4.90 & \textbf{5.00} &  7.23 &  5.47 & 4.89 \\ %
      q&               & CAEN DT5751 & 100 &   400 &  40.6 & 0.86 & 2.73 & \textbf{2.72} &  3.94 &  2.82 & 2.64 \\ %
\midrule
      r&                           & all values 0   &   1 &  1000 &  0   & 0    &  -   & \textbf{1.51} &  0.28 &  0.67 & 0.26 \\ %
      s& Flat traces $^\mathit{b}$  & all values 10  &   1 &  1000 &   0   & 0    &  -   & \textbf{1.51} &  0.28 &  0.67 & 0.26 \\ %
      t&                           & all values 100 &   1 &  1000 &  0   & 0    &  -   & \textbf{1.51} &  0.28 &  0.67 & 0.26 \\ %
\bottomrule
    \end{tabular}\par
    \bigskip
  }
  In \cref{tab:realtraces},
  the noise content of each trace collection is characterised by
  $\sigma_\mathcal{N}$, which is calculated from the distribution of
  the differences between each sample, and the average of its four
  closest neighbours on each side.
  The pulse amplitudes $A_\mathcal{P}$ are represented as a geometric average of the
  difference between the largest and smallest sample value in each
  trace.
  This slightly overestimates the pulse amplitudes, due to the noise
  broadening, but since $A_\mathcal{P} \gg \sigma_\mathcal{N}$, it
  is still clear that the traces contain pulses.
  The expected costs for storing the noise, as suggested by
  \cref{fig:comprsample}, are calculated as
  $\avgbr{c_\mathcal{S}} = \log_2 \sigma_\mathcal{N} + 2.95$.
  The LaBr$_3$ collections marked $^\mathit{a}$ are very flat
  (virtually no noise) except for the pulses, causing the DPTC
  costs to be dominated by its minimum of at least 1.5 bits/sample.
  This also applies to the synthetic constant-value traces marked
  $^\mathit{b}$.
\end{table*}

\subsection{Storage cost verification---actual traces}
\label{sec:compr_rate}

  \cref{tab:realtraces} shows the compression efficiencies for some different
  collections of actual data.  They are compared to the common gzip~\cite{web:gzip} and xz~\cite{web:xz} generic
  compression routines (at their normal setting).
  For the generic routines, all data of each file was stored in a
  binary file with 16-bit values.
  For a fair comparison, the overhead size of storing an empty compressed
  file was subtracted.
  In general, the DPTC results are quite similar to %
  the LZMA results, and well below the gzip results.
  The main exception are the LaBr$_3$ collections marked $^\mathit{a}$, where
  the data is very flat (virtually no noise) except for the pulses.
  Here the DPTC routine still uses its minimum of at least 1.5 bits/sample.
  This effect is also seen for the three synthetic traces marked $^\mathit{b}$,
  which have constant values.

  \revlabel{revrealhuffman}
  Since Huffman encoding \cite{art:huffman} is a common approach for compression where
  the typical distribution of values is known, the actual traces have
  for comparison purposes
  also been compressed using this approach.  It is applied after a
  difference stage, with the Huffman encodings individually optimised
  for each data set.  To allow average costs below one bit per sample
  for very flat traces, encodings of up to four consecutive values using
  one symbol were also allowed, when such stretches of values would
  account for more than \SI{1}{\percent} of the symbols.  In these
  tests, the \SI{1}{\percent} threshold was only passed for the cases
  marked $^\mathit{a}$ and $^\mathit{b}$.  Overall, the Huffman
  compression scheme delivers results slightly better than both the
  DPTC routine and the generic compression routines, but needs to be
  optimised to the characteristics of the signals.

  Finally, note how close the costs per sample are to the expectations for only
  storing the respective noise content, showing that the storage cost
  contributions from pulses are negligible.

\subsection{Caveat emptor---how to ignore ADC noise}

In case the original data contains an excessive number of least-significant bits with noise that shall not be stored, they must be shifted out of the original data before the values are given to the DPTC compression routine.
Just masking them out will \emph{not} improve the compression efficiency, as the routine is looking for the most significant bit of the differences that need to be stored.
On the other hand, using a compressor with $n$ larger than necessary causes little extra cost.
Few, if any, extra bits will be used; since mainly $k$ will potentially be affected, see \cref{fig:bitschunk}.

\revlabel{revkeepnoisybits}
Note that the choice of omitting least-significant bits is delicate decision.
The finally achievable resolution of a measurement may be
improved by retaining some additional least-significant bits, since it may
allow analysis of the later de-compressed traces to partially recover
the effects of quantization error and differential non-linearity in
the ADC, by averaging or fitting.

When applicable, in oversampled
parts of a trace, much larger savings than obtained through
omitting some least-significant bit may be obtained through
downsampling the information by summing adjacent samples before
compression,
thus storing fewer samples, but with better resolution.

\section{Conclusion}

A lossless compression routine which addresses both the transmission bandwidth and storage cost challenges associated with recording flash-ADC traces has been presented.
The routine can be directly integrated in front-end electronics and
\revlabel{revcanhandlefast} can
handle data streams on-the-fly at rates of \SI{400}{\mega samples/s} in the controlling FPGA.
Calculation of the differences between consecutive trace samples concentrated the most frequently occuring values around zero.
The compression was concluded by storing the values in groups of four, yielding a simple yet effective
variable-length code, by only storing the necessary least-significant bits, in a stream of 32-bit words.

A model for the storage cost was developed, by first considering the influence of the group headers as well as the retained ADC noise.  The additional cost of storing a pulse was expressed in terms of its amplitude and width.
By compressing a large set of artificial traces with varying characteristics, both the free parameters and the validity of the model were determined.

The method was then applied to actual data from different kinds of detectors.
The compression efficiency was found to be comparable to popular general-purpose compression methods (gzip and xz).
It was shown that the dominating cost of storing actual traces is generally given by
the retained ADC noise, and not the pulses.
It is therefore important for users to carefully assess how many least-significant bits shall be kept, in case they are noisy.
Except for that, there are no parameters that need to be adapted, which is of particular interest for experiments employing hundreds or thousands of detector channels.

Computer code for the FPGA implementation in VHDL and for the CPU decompression routine in C are available for download~\cite{web:dtpc} as open source software.

\section*{Acknowledgments}
\label{sec:Acknowledgment}
\addcontentsline{toc}{section}{Acknowledgment}

The authors would like to extend their thanks to
O. Schulz, B. L\"{o}her, S. Storck, and P. D\'{i}az Fern\'{a}ndez for providing test data,
and to A. Heinz and D. Radford for valuable discussions.

\bibliographystyle{IEEEtran}
\bibliography{IEEEabrv,dptc-article.bib}

\vfill

\end{document}